# Interfacial Dzyaloshinskii-Moriya interaction and spin-orbit torque in Au$_{1-x}$Pt$_x$/Co bilayers with varying interfacial spin-orbit coupling

Lijun Zhu[1,2,*], Lujun Zhu[3], Xin Ma[4], Xiaoqin Li[4], Robert A. Buhrman[1]

1. Cornell University, Ithaca, New York 14850, USA
2. State Key Laboratory of Superlattices and Microstructures, Institute of Semiconductors, Chinese Academy of Sciences, P.O. Box 912, Beijing 100083, China
3. College of Physics and Information Technology, Shaanxi Normal University, Xi'an 710062, China
4. Department of Physics, Center for Complex Quantum Systems, The University of Texas at Austin, Austin, Texas 78712, USA

E-mail: lz442@cornell.edu

The quantitative roles of the interfacial spin-orbit coupling (SOC) in Dzyaloshinskii-Moriya interaction (DMI) and dampinglike spin-orbit torque ($\tau_{DL}$) have remained unsettled after a decade of intensive study. Here, we report a conclusive experiment evidence that, because of the critical role of the interfacial orbital hybridization, the interfacial DMI is *not* necessarily a linear function of the interfacial SOC, e.g. at Au$_{1-x}$Pt$_x$/Co interfaces where the interfacial SOC can be tuned significantly via strongly composition ($x$)-dependent spin-orbit proximity effect without varying the bulk SOC and the electronegativity of the Au$_{1-x}$Pt$_x$ layer. We also find that $\tau_{DL}$ in the Au$_{1-x}$Pt$_x$/Co bilayers varies distinctly from the interfacial SOC as a function of $x$, indicating no important $\tau_{DL}$ contribution from the interfacial Rashba-Edelstein effect.

Spin-orbit coupling (SOC) phenomena is a central theme in condensed-matter physics. The SOC-induced Dzyaloshinskii-Moriya interaction (DMI) and spin-orbit torques (SOTs) in heavy metal/ferromagnet (HM/FM) bilayers have become two of the foundational aspects of spintronics research [1-7]. The DMI at HM/FM interfaces is a short-range anti-symmetric exchange interaction due to interfacial SOC and inversion asymmetry [1-3]. A strong interfacial DMI can compete with ferromagnetic exchange interaction and perpendicular magnetic anisotropy such that skyrmions [1,2] or Neél domain walls [3-5] can be stabilized and be displaced by SOT in chiral magnetic memories and logic. The interfacial DMI can affect micromagnetic non-uniformity [8] and thus magnetic damping [9], and ultrafast dynamics [10,11] during SOT switching of in-plane magnetization. The interfacial DMI also requires an in-plane magnetic field or its equivalent for SOT switching of perpendicular magnetization [12]. Despite the great technological importance and the intensive studies [13-21], the understanding of the underlying physics of the interfacial DMI has remained far from complete. Experiments have reported that the magnitude and in some cases even the sign of the interfacial DMI are phenomenally sensitive to the types of HM and FM [20], the HM thickness [19], or even atomic inter-diffusion at the interface [15,16,19]. However, despite the widespread recognition of its interfacial SOC nature [14-22], there has been no direct quantification of the interfacial DMI as a function of the strengths of the interfacial SOC ($\xi$) and the interfacial orbital hybridization in a HM/FM system. The basic question as the relative role and interplay of these two effects on the DMI has yet to be answered. A major challenge for any such experiments is how to widely vary and quantify $\xi$ in a HM/FM system that also exhibits a strong, accurately measurable DMI.

At the same time, the long-standing debate over the quantitative role of the interfacial SOC in the generation of the dampinglike SOT ($\tau_{DL}$) in HM/FM bilayers has remained unresolved. Some theories [22-25] and experiments [26-28] have indicated that the interfacial SOC can generate only a negligible $\tau_{DL}$ via the "two-dimensional" Rashba-Edelstein effect at HM/FM interfaces but instead a substantial loss of the spin angular momentum of an incident spin current to the lattice via interfacial spin-flip scattering [29,30]. In sharp contrast, other theories [31-33] predict that, if carriers can be scattered across the HM/FM interface, the $\tau_{DL}$ generated by interfacial SOC that can be comparable to that arising from the spin Hall effect (SHE) of the HM. The conclusions from experimental studies of the interfacial Rashba-Edelstein effect disagree strongly with regard to both the sign and magnitude of the associated $\tau_{DL}$ [7,34,35]. These theoretical and experimental divergences provide a strong motivation for a direct experimental determination of how $\tau_{DL}$ in a HM/FM sample is correlated with the interfacial SOC.

In this Letter, we demonstrate that the interfacial SOC at the Au$_{1-x}$Pt$_x$/Co interface can be tuned significantly via a strongly composition-dependent spin-orbit proximity effect (SOPE). From this ability, we establish that neither the interfacial DMI nor $\tau_{DL}$ of the Au$_{1-x}$Pt$_x$/Co heterostructure is a linear function of interfacial SOC due to the fundamental role of interfacial orbital hybridization and the absence of any significant interfacial $\tau_{DL}$, respectively.

For this study, magnetic bilayers of Au$_{1-x}$Pt$_x$ 4 nm/Co 2-7 nm ($x$ = 0, 0.25. 0.5, 0.65, 0.75, 0.85, and 1) were sputter-deposited onto oxidized Si substrates and capped with MgO 2 nm/Ta 1.5 nm protective layers (see more details in the Supplementary Materials [36]). The Au$_{1-x}$Pt$_x$/Co bilayers are textured with a face-centered-cubic (111) normal orientation [37]. The bilayers were patterned into 10×20 μm$^2$ microstrips for measuring the interfacial perpendicular magnetic anisotropy energy density ($K_s^{ISOC}$) and $\tau_{DL}$ via spin-torque ferromagnetic resonance (ST-FMR)[38,39].

It has been well established that $K_s^{ISOC}$ at the HM/Co interfaces originates from SOC-enhanced perpendicular orbital magnetic moments ($m_o^\perp$) localized at the first Co atomic layer adjacent to the interface [40,41]. According to Bruno's model [41,42], $K_s^{ISOC}/t_{FM} \propto \xi\ (m_o^\perp - m_o^\parallel)$, where $t_{FM}$ and $m_o^\parallel$ are the layer thickness and the in-plane orbital magnetic moment of the FM layer. Because $m_o^\perp = m_{o,i}^\perp/t_{FM} + m_o^\parallel$ [40,41], where $m_{o,i}^\perp$ is the $m_o^\perp$ value for the single FM interface layer and $m_o^\parallel$ reasonably approximates the bulk orbital magnetic moment value for the Co (0.13 ± 0.02 $\mu_B$/Co [40,41,43,44]), we obtain $K_s^{ISOC} \propto \xi m_{o,i}^\perp$ for the HM/Co interfaces. Previous experiments have determined that $m_{o,i}^\perp$ for the Au/Co interface (≈ 0.36 $\mu_B$/Co [41]) is twofold of that for the Pt/Co interface (≈ 0.18 $\mu_B$/Co [40,43]).

To quantify $K_s^{ISOC}$ for the $Au_{1-x}Pt_x$/Co interface using ST-FMR [36], we first determined the total interfacial perpendicular magnetic anisotropy energy density ($K_s$) of the two Co interfaces of the $Au_{1-x}Pt_x$/Co/MgO samples from the fits of the effective demagnetization field ($4\pi M_{eff}$) vs $t_{Co}^{-1}$ [Fig. 1(a)] to the relation $4\pi M_{eff} \approx 4\pi M_s + 2K_s/M_s t_{Co}$. We obtain $M_s \approx$ 1200-1300 emu/cm$^3$ and $K_s \approx$ 1.3-2.1 erg/cm$^2$ [Fig. 1(b)] for $Au_{1-x}Pt_x$/Co/MgO samples with different $x$. Using $K_s^{Co/MgO} \approx$ 0.32 erg/cm$^2$ as determined from a control sample of MgO/Co/MgO [36], we estimate $K_s^{ISOC}$ for the $Au_{1-x}Pt_x$/Co interfaces in Fig. 1(b). $K_s^{ISOC}$ increases from 1.02 erg/cm$^2$ for $x = 0$ to 1.69 erg/cm$^2$ for $x =$ 0.25-0.75, and then gradually decreases to 0.95 erg/cm$^2$ for $x = 1$. From this we can obtain $\xi_{Pt/Co}/\xi_{Au/Co} \approx 2$. Provided that $m_{orb,i}^\perp$ for the $Au_{1-x}Pt_x$/Co interfaces is not significantly smaller than that of the Pt/Co interface, the 1.8 times variation of $K_s^{ISOC}$ with $x$ would indicate a strong, but less than 3.6 times, tuning of $\xi$.

It is interesting to note that, despite the strong tuning of the interfacial SOC, the bulk SOC strength for the $Au_{1-x}Pt_x$ layer, which should vary in between that of pure Au and pure Pt [45], is expected to be approximately invariant with $x$ because Au and Pt have almost the same SOC strength (0.41 eV)[46]. We have also previously observed a threefold enhancement in the interfacial SOC of Pt/Co heterostructures by thermal engineering of the SOPE, without varying the composition and thus the bulk SOC of the HM [30]. These observations consistently demonstrate the distinct difference between the interfacial and the bulk SOCs, with the former being very sensitive to the local details of the interfaces.

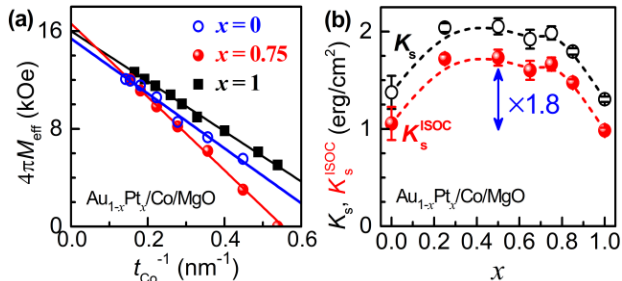

Fig. 1. (a) The $t_{Co}^{-1}$ dependence of $4\pi M_{eff}$ ($x = 1$, 0.75, and 0.65) of the $Au_{1-x}Pt_x$/Co bilayers. The some error bars are smaller than the points. The solid lines present the best linear fits. (b) Interfacial magnetic anisotropy energy density for the $Au_{1-x}Pt_x$/Co interface ($K_s^{ISOC}$) and for the two interfaces of $Au_{1-x}Pt_x$/Co/MgO ($K_s$) plotted as a function of $x$. The arrow indicates a factor of 1.8 variation of $K_s^{ISOC}$.

We determined the interfacial DMI of the $Au_{1-x}Pt_x$ 4 nm/Co 3.6 nm bilayers by measuring the DMI-induced frequency difference ($\Delta f_{DMI}$) between counter-propagating Damon-Eshbach spin waves using Brillouin light scattering (BLS)[13-21]. Figure 2(a) shows the geometry of the BLS measurements. The laser wavelength ($\lambda$) is 532 nm. The light incident angle ($\theta$) with respect to the film normal was varied from 0° to 32° to tune the magnon wave-vector ($k = 4\pi\sin\theta/\lambda$). A magnetic field ($H$) of ±1700 Oe was applied along the $x$ direction to align the magnetization of the Co layer. The anti-Stokes (Stokes) peaks in BLS spectra [Fig. 2(b)] correspond to the annihilation (creation) of magnons with $k$ (-$k$), while the total in-plane momentum is conserved during the BLS process. In Fig. 2(c), we plot $\Delta f_{DMI}$ as a function of $|k|$ for $Au_{1-x}Pt_x$/Co interface with different $x$. Here $\Delta f_{DMI}$ is the frequency difference of the ± $k$ peaks and averaged for $H = \pm1700$ Oe (see [20] for more details). The linear relation between $\Delta f_{DMI}$ and $k$ for each $x$ agrees with the expected relation [13,47] $\Delta f_{DMI} \approx (2\gamma/\pi\mu_0 M_s)Dk$, where $\gamma \approx$ 176 GHz/T is the gyromagnetic ratio and $D$ is the volumetric DMI constant.

As summarized in Fig. 2(d), with increasing $x$, $D$ for the $Au_{1-x}Pt_x$/Co interface varies by a factor of ~5, from -0.08 ± 0.02 for $x = 0$ to -0.45 ± 0.01 erg/cm$^2$ for $x = 0.85$. Taking into account the inverse dependence of $D$ on the FM thickness [16,17,19] due to the volume averaging effect, the total DMI strength for HM/FM interface can be estimated as $D_s = Dt_{FM}$. For the $Au_{1-x}Pt_x$/Co interfaces, $D_s$ changes gradually from $-30 \pm 6$ nerg/cm (1 nerg/cm =$10^{-9}$ erg/cm) at $x = 0$ to $-144 \pm 8$ nerg/cm at $x = 0.85$, and then drops slightly to $-114 \pm 6$ nerg/cm at $x = 1$. We first note that the DMI for the $Au_{1-x}Pt_x$/Co interfaces is very strong compared to those for other HM/FM systems. Even the relatively small $D_s \approx$ -30 nerg/cm for the Au/Co interface is already greater than all the reported values for (W, Ta, or Au)/(Co or FeCoB) interfaces ($|D_s| <$ 20 nerg/cm)[15,17,18,20]. The maximum value of $D_s \approx$ -144 nerg/cm for $Au_{0.85}Pt_{0.15}$/Co is comparable to the highest reported values for Pt/FeCoB (from -80 to -150 nerg/cm)[20,48,49], Pt/Co (< -160 nerg/cm)[19,20], and Ir/Co (from 30 to 140 nerg/cm)[20,21]. The large DMI amplitude and the fivefold tunability provided by the $Au_{1-x}Pt_x$ composition make $Au_{1-x}Pt_x$/Co an especially intriguing system for chiral spintronics.

The fivefold variation of interfacial DMI with $x$ for the $Au_{1-x}Pt_x$/Co interfaces is clearly not in linear proportion to the 1.8 times (<3.6 times) variation of $K_s^{ISOC}$ ($\xi$). Indeed when we plot $D$ as a function of $K_s^{ISOC}$ [Fig. 2(e)], it becomes quite apparent that there is not a linear correlation between $D$ and $K_s^{ISOC}$. This is a strikingly observation as it indicates that there must be another composition-sensitive effect that is, at least, as critical as the interfacial SOC for interfacial DMI. So far, effects including electronegativity [4], intermixing [14], proximity magnetism [14], orbital anisotropy [43], and orbital hybridization [14,20] have been suggested to affect DMI. Since Au and Pt have quite similar electronegativities (2.2 for Pt and 2.4 for Au)[50], a substantial composition-induced electronegativity variation seems unlikely to occur at the $Au_{1-x}Pt_x$/Co interface. Previous studies have suggested that interfacial alloying, if significant, may substantially degrade both the DMI [14] and $K_s^{ISOC}$ [30,51]. However, our $Au_{1-x}Pt_x$/Co bilayers have very high $D_s$ and $K_s^{ISOC}$. In previous work, we have reported on the structural characterizations of HM/FM samples prepared with similar conditions by transmission electron microscopy [52], x-ray reflectivity, and secondary ion mass spectrometry [30], all of which have indicated minimal interface alloying. The relatively low values of $M_s$ (1200-1300 emu/cm$^3$) also indicate a rather minimal proximity magnetism at these $Au_{1-x}Pt_x$/Co interfaces [53]. The DMI in Pt/Co bilayers was also suggested to correlate linearly with orbital anisotropy, with the latter being quantified by the $m_{o,i}^\perp/m_o^\parallel$ ratio [43]. Since $m_o^\parallel$ of Co is approximately invariant [40,41,43], we expect $K_s^{ISOC} \propto \xi m_{o,i}^\perp/m_o^\parallel$, due to which the combined effect of the orbital anisotropy and the interfacial SOC would be a linear dependence of $D$ on $K_s^{ISOC}$. Thus, from the lack of a linear correlation between $D$ and $K_s^{ISOC}$ in Fig. 2(e) we conclude that the variation of the DMI at the $Au_{1-x}Pt_x$/Co interfaces with $x$ cannot be attributed to the change of orbital anisotropy.

As we discuss below, the non-monotonic and exceptionally strong tunability of the DMI (stronger than that can be provided solely by the interfacial SOC) can be understood by taking into account the essential role of the varying orbital hybridization at the $Au_{1-x}Pt_x$/Co interface. Theoretical calculations have shown that, besides the interfacial SOC, the 3d orbital occupations and their spin-flip scattering with the spin-orbit active 5d states collectively control the overall DMI [54]. The strength of the interfacial orbital hybridization is expected to be inversely proportional to the on-site energy difference of the 3d and 5d states [55]. The Fermi level ($E_F$) of bulk Au (5d$^{10}$6s$^1$) is located ~2 eV above the top of 5d band so that there are no 5d orbitals at the Fermi surface [56]. In contrast, $E_F$ of bulk Pt (5d$^9$6s$^1$) is located in the top region of its 5d band. In both bulk Au and Pt [36], the density of states (DOS) has a sharp peak at the top region of the 5d band [56]. First-principles calculations [57] have indicated that, in magnetic multilayers consisting of repeats of HM (2 monolayers)/Co (monolayer) multilayers where the interfacial orbital hybridization becomes very important, both the HM 5d bands and the Co 3d bands are significantly broadened compared to their bulk properties. In Fig. 2(f), we compare the calculated results [57] for the local DOS of the Au/Co and Pt/Co systems. For Au/Co, the top of the Au 5d band is located ≈1.2 eV below $E_F$, the minority spin band of Co 3d is centered at $E_F$. For the Pt/Co system, the top of the Pt 5d band is 0.7 eV above $E_F$, and the minority spin band of Co 3d is 0.63 eV above $E_F$ for the Pt/Co. In both cases, the majority spin band of Co 3d is located well below $E_F$ due to the exchange splitting.

As $x$ increases, the 5d band of the $Au_{1-x}Pt_x$/Co interfaces can be expected to be lifted continuously and pass through the Fermi level (i.e. from -1.2 eV for $x = 0$ to +0.7 eV for $x = 1$ with respect to $E_F$). Meanwhile, the top of the minority spin band of Co 3d should also be shifted to be well above $E_F$. As a consequence, with increasing $x$, the hybridization of the 5d orbitals of the $Au_{1-x}Pt_x$ with the 3d orbitals of Co would be first strengthened (mainly due to the shift of 5d band), then be maximized at an intermediate composition where the DOS peak of the 5d band is approximately at $E_F$ and is well aligned with the DOS peak of the 3d minority spin band of Co, and then finally decrease slightly as $x$ approaches 1 (the DOS peak of the minority spin band Co 3d is moved away from $E_F$). This is quite consistent with our experimental observation that, as $x$ increases, the DMI is first increasingly enhanced by, we propose, 5d-3d hybridization, then is maximized at about $x = 0.85$, and finally decrease [Fig. 2(d)]. The moderate reduction of interfacial SOC for $x > 0.85$ [see $K_s^{ISOC}$ in Fig. 1(c)] is also a likely contributor to the decrease of the DMI in this composition range.

In our previous study of $Pd_{1-x}Pt_x$/$Fe_{60}Co_{20}B_{20}$ interfaces [48], we found that the DMI does vary proportionally with $K_s^{ISOC}$, which suggests a negligible variation of the interfacial orbital hybridization in that particular material system. This seems reasonable because, in the bulk, the energy distribution of the 4d DOS of Pd (4d$^{10}$5s$^0$) is rather analogous to that of Pt 5d, with the first DOS peak of the Pd 4d band lying at about $E_F$ [36,56]. Since the orbital hybridization at the $Pd_{1-x}Pt_x$/$Fe_{60}Co_{20}B_{20}$ interfaces can further broaden the 5d and 3d bands, the DOS distribution and thus the strength of the interfacial 5d-3d orbital hybridization should be reasonably similar as a function of the $Pd_{1-x}Pt_x$ composition. This conclusion is very well supported by a first-principles calculation [58]. Pd also has the same electronegativity (2.2) [50] and valence electron number (10) [46] as Pt. Therefore, in the case of $Pd_{1-x}Pt_x$/$Fe_{60}Co_{20}B_{20}$, the interfacial SOC is left as the only important variable that determines the variation of the interfacial DMI with the $Pd_{1-x}Pt_x$ composition.

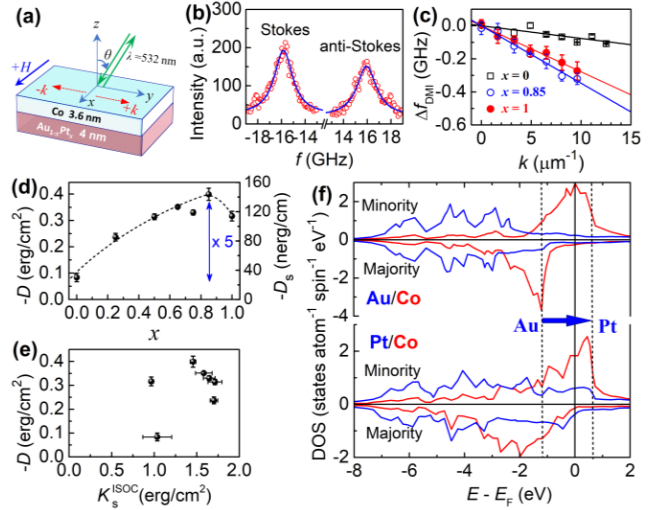

Fig. 2. (a) BLS measurement geometry; (b) BLS spectra ($k = 8.1$ μm$^{-1}$, $H = 1700$ Oe, $x = 0.85$), (c) $k$ dependence of $\Delta f_{DMI}$ ($x = 0$, 0.85, and 1), (d) $D$, $D_s$, and (e) $K_s^{ISOC}$ dependence of $D$ for $Au_{1-x}Pt_x$ 4 nm/Co 3.6 nm bilayers with different $x$. (f) Calculated local DOS for Au/Co (up) and Pt/Co (down) interfaces (blue for the DOS of Au and Pt, red for that of Co, the data are taken from [55]), highlighting a substantial shift of the 5d band from below to beyond $E_F$ as a function of $x$. The blue solid curves in (b) represent fits to the Lorentzian function. The solid lines in (c) refer to the linear fits. The two dashed lines in (f) refer to the top of the 5d bands of Au and Pt.

The SOTs due to an interfacial Rashba-Edelstein effect should increase linearly with the Rashba constant ($\alpha_R$) [22-25]. It has been established that $\alpha_R \propto \xi\, m_{o,i}^{\perp}/m_o^{\parallel}$ [44] and that $m_o^{\parallel}$ is approximately constant at the HM/Co interfaces [40,41,43,44]. As a result, $K_s^{ISOC}$ of the $Au_{1-x}Pt_x$/Co interfaces is a good linear indicator for $\alpha_R$. The strong tunability of $K_s^{ISOC}$ and thus $\alpha_R$ of the $Au_{1-x}Pt_x$/Co interfaces with $x$ provides a novel access to test for any significant $\tau_{DL}$ due to the Rashba-Edelstein effect. If we define the apparent FMR spin-torque efficiency ($\xi_{FMR}$) from the ratio of the symmetric and anti-symmetric components of the magnetoresistance response of the ST-FMR [38,39], the efficiency of $\tau_{DL}$ per current density ($\xi_{DL}^j$) for the $Au_{1-x}Pt_x$/Co bilayers can be determined as the inverse intercept in the linear fit of $\xi_{FMR}^{-1}$ vs $t_{Co}^{-1}$ [Fig. 3(a)]. In Fig. 3(b), we summarize the ST-FMR results of $\xi_{DL}^j$ for $Au_{1-x}Pt_x$ 4 nm/Co 2-7 nm together with the "in-plane" harmonic response results for $Au_{1-x}Pt_x$ 4 nm/Co 1.4 nm bilayers, "out-of-plane" harmonic response results for $Au_{1-x}Pt_x$ 4 nm/Co 0.8 nm bilayers [37], and SOT switching results for in-plane SOT-magnetic tunnel junctions (MTJs)[11] with a SHE channel of 4 nm $Au_{0.25}Pt_{0.75}$ and a magnetic free layer of 1.4 nm $Fe_{0.6}Co_{0.2}B_{0.2}$. Although the $\xi_{DL}^j$ values from ST-FMR are approximately threefold smaller than those from the other measurements (see more examples in [52]) for some reasons yet to resolve, the four different measurements are qualitatively quite consistent with respect to the functional dependence on $x$, with $\xi_{DL}^j$ increasing by 15 times when $x$ increases from 0 to 0.75. When we take into account the spin memory loss to the lattice by interfacial spin-flip scattering, which increases linearly with $K_s^{ISOC}$ [30], the variation in the spin current generation is much greater than a factor of 15. We also find that the fieldlike SOT of the $Au_{1-x}Pt_x$/Co samples is minimal and uncorrelated to $\alpha_R$ [36].

This result disagrees with any significant interfacial torques because the latter, if any, should be proportional to $\alpha_R$ or $K_s^{ISOC}$ so that it should vary by only a factor of 1.8. This conclusion is also supported by our recent experiments that the thermally-tuned interfacial SOC at HM/FM interfaces acts as spin memory loss rather than generates SOTs [30]. Instead, as discussed in [37] the variation of $\xi_{DL}^j$ can be well attributed to the composition-dependent resistivity and intrinsic spin Hall conductivity of the $Au_{1-x}Pt_x$ layer. It is also an interesting observation that $\xi_{DL}^j$ for $Au_{1-x}Pt_x$ 4/Co varies by a factor of 15 when the bulk SOC in $Au_{1-x}Pt_x$ remains the same. This is consistent with the intrinsic SHE being determined by the spin Berry curvature of the band structure of the HM rather than simply the bulk SOC strength [46]. Therefore, $\tau_{DL}$ of a HM/FM interfaces, at least in this system, behaves very differently from the interfacial Rashba-Edelstein effect, the interfacial DMI, and the bulk SOC.

In summary, we have demonstrated that the interfacial SOC at $Au_{1-x}Pt_x$/Co can be tuned significantly via a strongly composition-dependent SOPE, without varying the bulk SOC and the electronegativity of the HM. We find that both the interfacial SOC and the interfacial orbital hybridization of the HM/FM interfaces play critical roles in the determination of the DMI. As a consequence, the interfacial DMI is not always a linear indicator of interfacial SOC, e.g. in the $Au_{1-x}Pt_x$/Co bilayers where the interfacial orbital hybridization varies substantially with composition. The distinct composition dependences of interfacial SOC and $\tau_{DL}$ suggest minimal $\tau_{DL}$ from the Rashba-Edelstein effect. This is only consistent with the theoretical prediction that the localized Rashba-Edelstein effect contributes to negligible $\tau_{DL}$ [22-25]. These findings provide an in-depth understanding of interfacial SOC, interfacial DMI, and $\tau_{DL}$, which will advance the development of high-efficient chiral magnetic devices. The large amplitudes and the strong tunability of both the DMI and $\tau_{DL}$ provided by the $Au_{1-x}Pt_x$ composition make $Au_{1-x}Pt_x$/Co heterostructure an especially intriguing system for chiral spintronics.

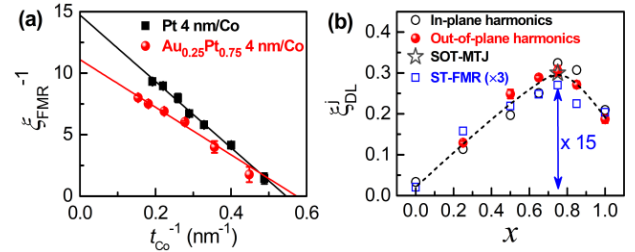

Fig. 3. (a) $\xi_{FMR}^{-1}$ vs $t_{Co}^{-1}$ for Pt/Co and $Au_{0.25}Pt_{0.75}$/Co bilayers (ST-FMR), (b) $x$ dependence of $\xi_{DL}^j$ for the $Au_{1-x}Pt_x$/Co bilayers as determined from in-plane harmonics, out-of-plane harmonics [37], SOT-MTJ [11], and ST-FMR measurements. The solid lines in (a) represent the linear fits to the data. In (b), the $\xi_{DL}^j$ values from the ST-FMR are multiplied by 3 for clarity, some error bars are smaller than the points. The blue arrow indicates a factor of 15 variation of $\xi_{DL}^j$, the dashed line is just for the guidance of the eyes.

We thank Kemal Sobotkiewich for collecting the BLS data. This work was supported in part by the Office of Naval Research (N00014-19-1-2143), in part by the NSF MRSEC program (DMR-1719875) through the Cornell Center for Materials Research, and in part by the Defense Advanced Research Projects Agency (USDI D18AC00009). The devices were fabricated in part at the Cornell NanoScale Facility, an NNCI member supported by NSF Grant ECCS-1542081. The DMI measurement performed at UT-Austin was primarily supported as part of SHINES, an Energy Frontier Research Center funded by the U.S. Department of Energy (DOE), Office of Science, Basic Energy Science (BES) under Award No. DE-SC0012670 and partially supported by the National Science Foundation through the Center for Dynamics and Control of Materials: an NSF MRSEC under Cooperative Agreement No. DMR1720595.